\documentclass[zpreprint,zbstpl]{LaTeX/zeus/zeus_paper}
\usepackage[english]{babel}
\newcommand{\Zdetdesc}{%
A detailed description of the ZEUS detector can be found 
elsewhere~\cite{zeus:1993:bluebook}. A brief outline of the 
components that are most relevant for this analysis is given
below.\xspace}

\chardef\usc=95
\chardef\til=126
\catcode`\@=11 
\DeclareRobustCommand\xdotspace{\futurelet\@let@token\@xdotspace}
\def\@xdotspace{%
  \ifx\@let@token.\else
  \ifx\@let@token\bgroup.\else
  \ifx\@let@token\egroup.\else
  \ifx\@let@token\/.\else
  \ifx\@let@token\ .\else
  \ifx\@let@token~.\else
  \ifx\@let@token!.\else
  \ifx\@let@token,.\else
  \ifx\@let@token:.\else
  \ifx\@let@token;.\else
  \ifx\@let@token?.\else
  \ifx\@let@token/.\else
  \ifx\@let@token'.\else
  \ifx\@let@token).\else
  \ifx\@let@token-.\else
  \ifx\@let@token\@xobeysp.\else
  \ifx\@let@token\space.\else
  \ifx\@let@token\@sptoken.\else
   .\space
   \fi\fi\fi\fi\fi\fi\fi\fi\fi\fi\fi\fi\fi\fi\fi\fi\fi\fi}
\catcode`\@=12 

\newcommand{\stru}[2]{%
   \relax\ifmmode\hbox{\vrule height#1 depth#2 width0pt}%
   \else\vrule height#1 depth#2 width0pt\fi}

\newcommand{\Ronum}[1]{\uppercase\expandafter{\romannumeral#1}}
\newcommand{\ronum}[1]{\expandafter{\romannumeral#1}}
\DeclareRobustCommand{\LaTeXZ}{%
  \LaTeX\kern-.05em4\kern-.1em
  {\raisebox{-0.2ex}{$\scriptstyle\text{ZEUS}$}}\xspace}

\newcommand{\fig}[1]{Fig.~\ref{fig-#1}}
\newcommand{\Fig}[1]{Figure~\ref{fig-#1}}

\newcommand{\tab}[1]{Table~\ref{tab-#1}}

\newcommand{\Sect}[1]{Section~\ref{sec-#1}}

\DeclareMathAlphabet{\mathbf}{OT1}{cmr}{bx}{sl}
\newcommand{\eVdist}{\kern-0.06667em}

\newcommand{\Gev}{{\text{Ge}\eVdist\text{V\/}}}

\newcommand{\gev}{{\,\text{Ge}\eVdist\text{V\/}}}

\newcommand{\pbi}{\,\text{pb}^{-1}}

\newcommand{\met}{\,\text{m}}

\newcommand{\cm}{\,\text{cm}}

\newcommand{\rad}{\,\text{rad}}

\newcommand{\Tesla}{\,\text{T}}

\newcommand{\slashfrac}[2]{%
  \raisebox{0.5ex}{\ensuremath #1}\kern-0.12em/\kern-0.08em
  \raisebox{-.8ex}{\ensuremath #2}}

\newcommand{\sqr}[3]{%
    {\vcenter{\hrule height.#3ex\hbox{\vrule width.#2ex height#1ex
     \kern#1ex\vrule width.#3ex}\hrule height.#2ex}}}

\catcode`\@=11 
\newcommand{\parenbar}{\mathpalette\p@renb@r}
\def\p@renb@r#1#2{\vbox{%
  \ifx#1\scriptscriptstyle \dimen@.7em\dimen@ii.2em\else
  \ifx#1\scriptstyle \dimen@.8em\dimen@ii.25em\else
  \dimen@1em\dimen@ii.4em\fi\fi \offinterlineskip
  \ialign{\hfill##\hfill\cr
    \vbox{\hrule width\dimen@ii}\cr
    \noalign{\vskip-.3ex}%
    \hbox to\dimen@{$\mathchar300\hfil\mathchar301$}\cr
    \noalign{\vskip-.3ex}%
    $#1#2$\cr}}}
\catcode`\@=12 

\newcommand{\IP}{{\rm I$\kern-0.01667em$P}\xspace}

\mathchardef\qsm=63
\mathchardef\pls=43
\mathchardef\mns=512
\mathchardef\plm=518
\mathchardef\eql=61
\mathchardef\smallleft=300
\mathchardef\smallright=301
\mathchardef\les=316
\mathchardef\gre=318
\mathchardef\leq=532
\mathchardef\grq=533

\catcode`\@=11 
\newcounter{pict@width}
\newcounter{pict@height}
\newlength{\pict@scale}
\newcommand{\psfigadd}[4]{%
\setcounter{pict@width}{1*\ratio{#2+\pict@scale/2}{\pict@scale}}
\setcounter{pict@height}{1*\ratio{#3+\pict@scale/2}{\pict@scale}}
\setlength{\unitlength}{\pict@scale}
\hbox to #2{\hspace{-\fill}\begin{picture}(\thepict@width,\thepict@height)
\put(0,0){\psfig{figure=#1,width=#2,height=#3,clip=}}
\SetScale{0.283466457}
\SetWidth{1.763889}
{#4}
\end{picture}}
}
\newcounter{pict@widthfst}
\newcounter{pict@widthscd}
\newcounter{pict@widthtot}
\newcommand{\psfigaddtwo}[7]{%
\setcounter{pict@widthfst}{1*\ratio{#2+\pict@scale/2}{\pict@scale}}
\setcounter{pict@widthscd}{1*\ratio{#2+#4+\pict@scale/2}{\pict@scale}}
\setcounter{pict@widthtot}{1*\ratio{#2+#4+#6+\pict@scale/2}{\pict@scale}}
\setcounter{pict@height}{1*\ratio{#3+\pict@scale/2}{\pict@scale}}
\setlength{\unitlength}{\pict@scale}
\hbox{\hspace{-\fill}\begin{picture}(\thepict@widthtot,\thepict@height)
\put(0,0){\psfig{figure=#1,width=#2,height=#3,clip=}}
\put(\thepict@widthscd,0){\psfig{figure=#5,width=#6,height=#3,clip=}}
\SetScale{0.283466457}
\SetWidth{1.763889}
{#7}
\end{picture}}
}
\newcommand{\psfigror}[4]{%
\setcounter{pict@width}{1*\ratio{#2+\pict@scale/2}{\pict@scale}}
\setcounter{pict@height}{1*\ratio{#3+\pict@scale/2}{\pict@scale}}
\setlength{\unitlength}{\pict@scale}
\hbox{\begin{picture}(\thepict@width,\thepict@height)
\put(0,\thepict@height){\psfig{figure=#1,width=#3,height=#2,clip=,angle=270}}
\SetScale{0.283466457}
\SetWidth{1.763889}
{#4}
\end{picture}}
}
\newcommand{\psfigrol}[4]{%
\setcounter{pict@width}{1*\ratio{#2+\pict@scale/2}{\pict@scale}}
\setcounter{pict@height}{1*\ratio{#3+\pict@scale/2}{\pict@scale}}
\setlength{\unitlength}{\pict@scale}
\hbox{\begin{picture}(\thepict@width,\thepict@height)
\put(0,0){\psfig{figure=#1,width=#3,height=#2,clip=,angle=90}}
\SetScale{0.283466457}
\SetWidth{1.763889}
{#4}
\end{picture}}
}
\catcode`\@=12 
\newlength\listtextwidth

\catcode`\@=11 
\newlength{\@tabfninsert}
\newlength{\@tabfnwidth}
\newcommand{\tabfootnote}[2]{%
  \setlength{\@tabfninsert}{0.8em}
  \setlength{\@tabfnwidth}{\textwidth}
  \addtolength{\@tabfnwidth}{-\@tabfninsert}
  \addtolength{\@tabfnwidth}{-0.4em}
  \noindent\makebox[\@tabfninsert][r]{\footnotesize$^{#1}$\hfil}\hfill%
  \parbox[t]{\@tabfnwidth}{\footnotesize #2\hfill}}
\catcode`\@=12 

\newcommand{\ptmiss}{\not\hspace{-0.55ex}{P}_t}

\newcommand{\rp}{$\not$\kern-0.pt$R_p$}

\def\citeCTD{{\cite{%
nim:a279:290,*npps:b32:181,*nim:a338:254%
}}\xspace}
\def\citeCAL{{\cite{%
nim:a309:77,*nim:a309:101,*nim:a321:356,*nim:a336:23%
}}\xspace}

\begin{document}
\prepnum{DESY--01--132} 

\title{
Searches for excited fermions\\
in {\boldmath $ep$} collisions at HERA
}
                    
\author{ZEUS Collaboration}
\date{July 2002}

\abstract{ Searches in $ep$ collisions 
for heavy excited fermions have been performed 
with the ZEUS detector at HERA. Excited states of electrons and
quarks have been searched for in $e^+p$ collisions at a centre-of-mass energy of
$300\gev$ using an integrated luminosity of $47.7\pbi$. Excited electrons have
been sought via the decays $e^* \to e \gamma$, $e^* \to e Z$ and $e^* \to \nu
W$. Excited quarks have been sought via the decays $q^* \to q \gamma$ and $q^*
\to q W$. A search for excited neutrinos decaying via $\nu^*\to \nu\gamma$,
$\nu^* \to \nu Z$ and $\nu^* \to e W$ is presented using $e^-p$ collisions at
$318\gev$ centre-of-mass energy, corresponding to an integrated luminosity of
$16.7\pbi$.  No evidence for any excited fermion is found, and limits on the
characteristic couplings are derived for masses $\lesssim 250\gev$.  }

\makezeustitle

%
%
%
%
\def\3{\ss}   
\newcommand{\address}{ }                                                                           
\pagenumbering{Roman}                                                                              

\begin{center}                                                                                     
{                      \Large  The ZEUS Collaboration              }                               
\end{center}                                                                                       
  S.~Chekanov,                                                                                     
  M.~Derrick,                                                                                      
  D.~Krakauer,                                                                                     
  S.~Magill,                                                                                       
  B.~Musgrave,                                                                                     
  A.~Pellegrino,                                                                                   
  J.~Repond,                                                                                       
  R.~Yoshida\\                                                                                     
 {\it Argonne National Laboratory, Argonne, Illinois 60439-4815}~$^{n}$                            
\par \filbreak                                                                                     
  M.C.K.~Mattingly \\                                                                              
 {\it Andrews University, Berrien Springs, Michigan 49104-0380}                                    
\par \filbreak                                                                                     
  P.~Antonioli,                                                                                    
  G.~Bari,                                                                                         
  M.~Basile,                                                                                       
  L.~Bellagamba,                                                                                   
  D.~Boscherini$^{   1}$,                                                                          
  A.~Bruni,                                                                                        
  G.~Bruni,                                                                                        
  G.~Cara~Romeo,                                                                                   
  L.~Cifarelli$^{   2}$,                                                                           
  F.~Cindolo,                                                                                      
  A.~Contin,                                                                                       
  M.~Corradi,                                                                                      
  S.~De~Pasquale,                                                                                  
  P.~Giusti,                                                                                       
  G.~Iacobucci,                                                                                    
  G.~Levi,                                                                                         
  A.~Margotti,                                                                                     
  T.~Massam,                                                                                       
  R.~Nania,                                                                                        
  F.~Palmonari,                                                                                    
  A.~Pesci,                                                                                        
  G.~Sartorelli,                                                                                   
  A.~Zichichi  \\                                                                                  
  {\it University and INFN Bologna, Bologna, Italy}~$^{e}$
% ...
\newpage                                                                                           
$^{\    1}$ now visiting scientist at DESY \\                                                      
$^{\    2}$ now at Univ. of Salerno and INFN Napoli, Italy \\                                      
% ...
                                                           %
                                                           %
\newpage   
                                                           %
                                                           %
\begin{tabular}[h]{rp{14cm}}                                                                       
$^{a}$ &  supported by the Natural Sciences and Engineering Research                               
          Council of Canada (NSERC)  \\                                                            
$^{b}$ &  supported by the German Federal Ministry for Education and                               
          Research (BMBF), under contract numbers HZ1GUA 2, HZ1GUB 0,                              
          HZ1PDA 5, HZ1VFA 5\\                                                                     
$^{c}$ &  supported by the MINERVA Gesellschaft f\"ur Forschung GmbH, the                          
          Israel Science Foundation, the U.S.-Israel Binational Science                            
          Foundation, the Israel Ministry of Science and the Benozyio Center                       
          for High Energy Physics\\                                                                
$^{d}$ &  supported by the German-Israeli Foundation, the Israel Science                           
          Foundation, and by the Israel Ministry of Science \\                                     
$^{e}$ &  supported by the Italian National Institute for Nuclear Physics                          
          (INFN) \\                                                                                
$^{f}$ &  supported by the Japanese Ministry of Education, Science and                             
          Culture (the Monbusho) and its grants for Scientific Research \\                         
$^{g}$ &  supported by the Korean Ministry of Education and Korea Science                          
          and Engineering Foundation  \\                                                           
$^{h}$ &  supported by the Netherlands Foundation for Research on                                  
          Matter (FOM) \\                                                                          
$^{i}$ &  supported by the Polish State Committee for Scientific Research,                         
          grant no. 2P03B04616, 620/E-77/SPUB-M/DESY/P-03/DZ 247/2000-2002                         
          and 112/E-356/SPUB-M/DESY/P-03/DZ 3001/2000-2002\\                                       
$^{j}$ &  partially supported by the German Federal Ministry for Education                         
          and Research (BMBF)  \\                                                                  
$^{k}$ &  supported by the Fund for Fundamental Research of Russian Ministry                       
          for Science and Edu\-cation and by the German Federal Ministry for                       
          Education and Research(BMBF) \\                                                          
$^{l}$ &  supported by the Spanish Ministry of Education                                           
          and Science through funds provided by CICYT \\                                           
$^{m}$ &  supported by the Particle Physics and                                                    
          Astronomy Research Council, UK \\                                                        
$^{n}$ &  supported by the US Department of Energy \\                                              
$^{o}$ &  supported by the US National Science Foundation                                          
\end{tabular}                                                                                      
                                                           %

\pagenumbering{arabic} 
\pagestyle{plain}
\section{Introduction}
\label{sec-int}

\begin{figure}[h]

    \begin{minipage}{0.3\textwidth}
      \epsfig{file=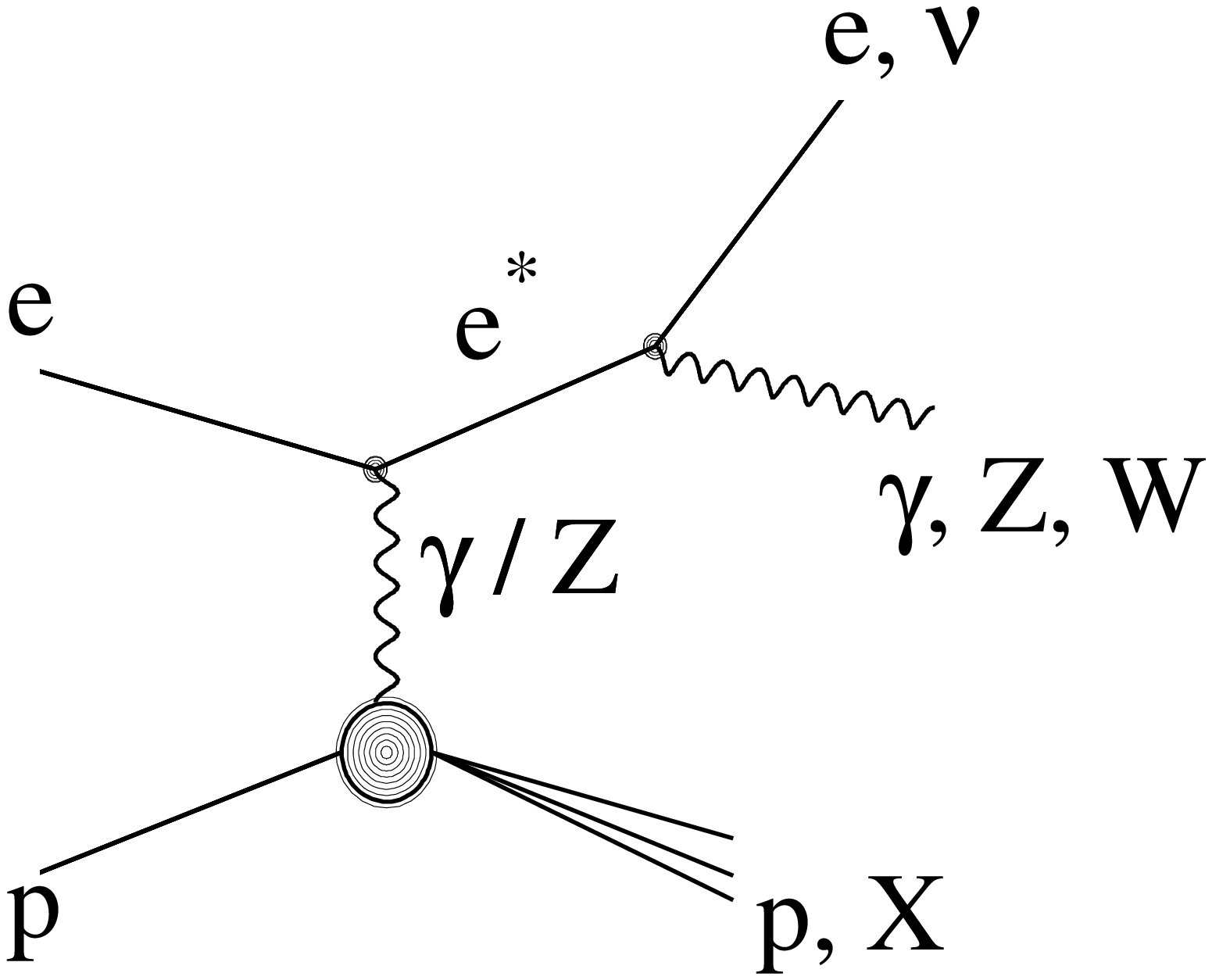,height=4cm}\\a)
    \end{minipage}\hfill
    \begin{minipage}{0.3\textwidth}
      \epsfig{file=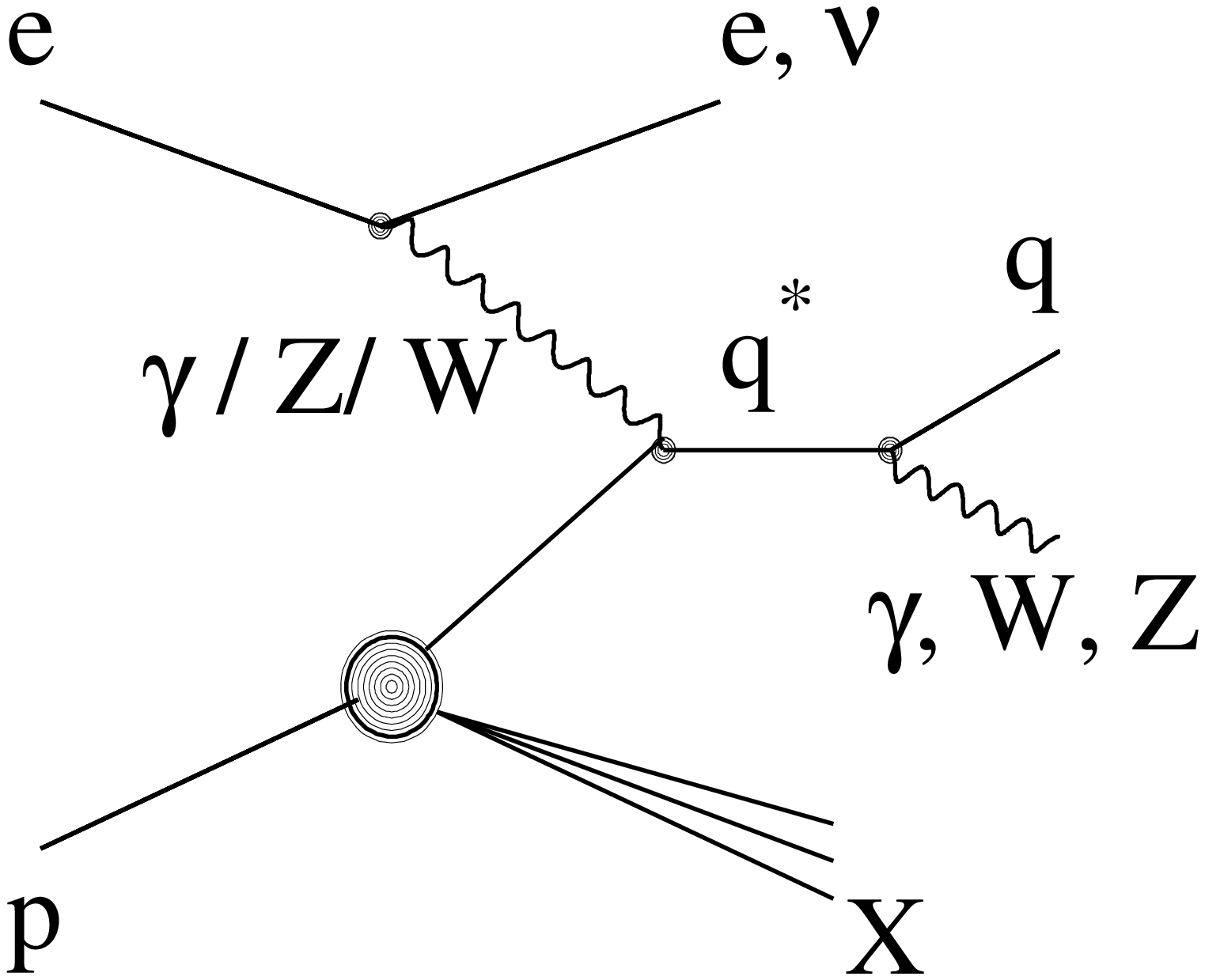,height=4cm}\\b)
    \end{minipage}\hfill
    \begin{minipage}{0.3\textwidth}
      \epsfig{file=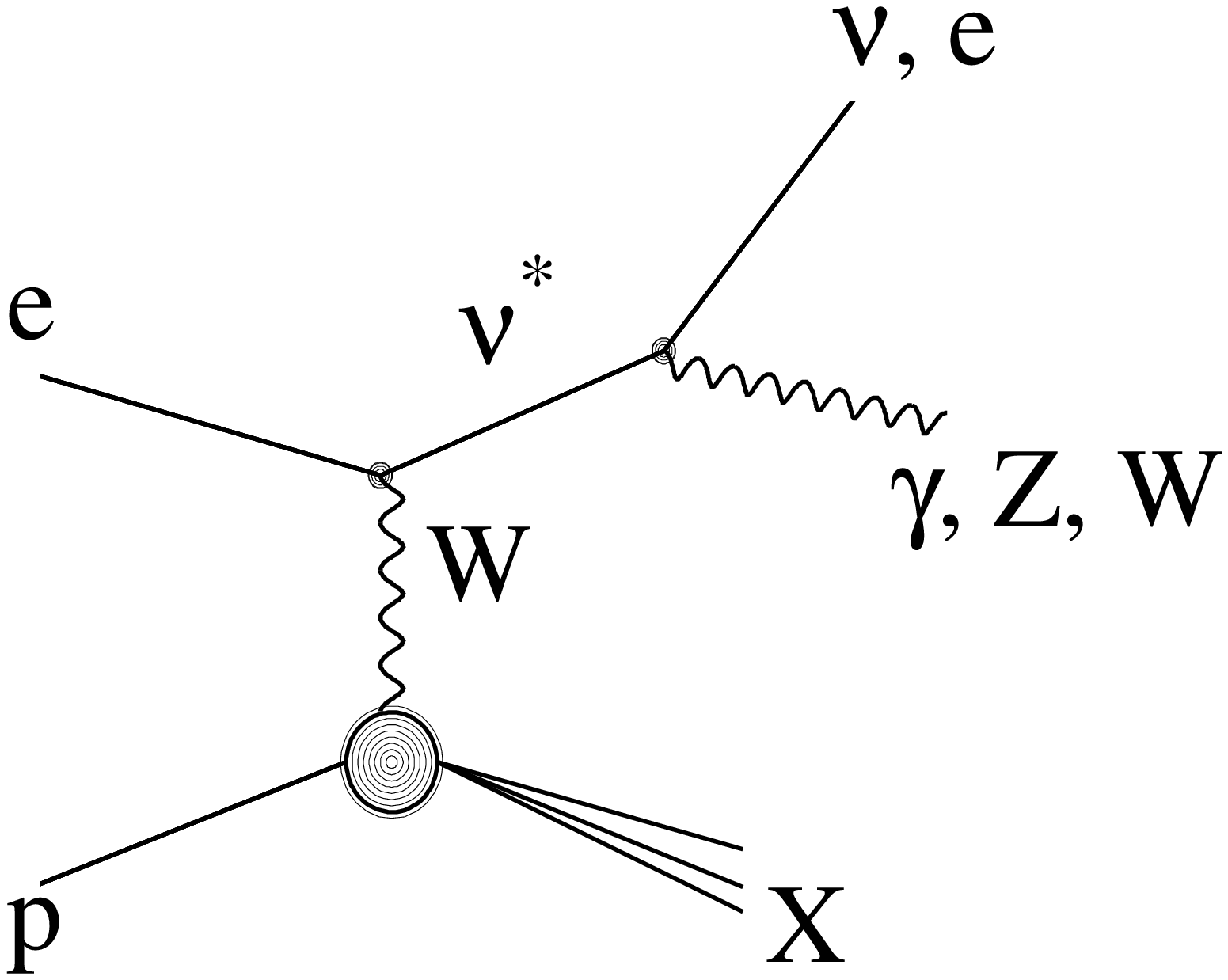,height=4cm}\\c)
    \end{minipage}

  \caption{Diagrams considered for the production of (a) excited
    electrons, (b) excited quarks and (c) excited neutrinos in $ep$ collisions,
    with their decays into a known fermion and a gauge boson.}

  \label{fig-feynman}
\end{figure}

The large number of quarks and leptons in the Standard Model suggests the
possibility that they may be composite particles, consisting of combinations of
more fundamental entities.
The observation of excited states of quarks or leptons would be a clear signal
that these particles are composite rather than elementary.  At the
electron\footnote{Throughout this paper, ``electron'' is used generically to
  refer to $e^+$ as well as $e^-$.}-proton collider HERA, excited electrons,
quarks and neutrinos ($e^*$, $q^*$, $\nu^*$) with masses up to the kinematic
limit of $318\gev$ could be produced directly via $t$-channel exchange of a
gauge boson as shown in Fig.~\ref{fig-feynman}: for $e^*$, via $\gamma/Z$
exchange; for $q^*$, via $\gamma/Z/W$ exchange; and for $\nu^*$,
 via $W$ exchange. Once produced, the excited fermion ($F^*$) decays 
into a known fermion and a gauge boson.

This paper reports on searches for excited electrons and quarks in $e^+p$
collisions and for excited neutrinos in $e^-p$ collisions at HERA.  From 1994 to
1997, the HERA collider operated with positron and proton energies of $27.5\gev$
and $820\gev$, respectively, resulting in a centre-of-mass energy of $300\gev$.
A total of $47.7\pbi$ of data were collected in the ZEUS detector during this
period. This corresponds to a five-fold increase in statistics over the
previously published ZEUS search with $e^+p$ data~\cite{zfp:c76:631}. A search
for excited fermions based on $37\pbi$ of $e^+p$ data has been reported recently
by the H1 collaboration~\cite{epj:c17:567}. In 1998 and 1999, the collider
operated with $e^-$ and with an increased proton energy of $920\gev$, resulting
in a centre-of-mass energy of $318\gev$.  The data collected with the ZEUS
detector during this period correspond to an integrated luminosity of
$16.7\pbi$, leading to a 30-fold increase in statistics over the previous ZEUS
publication with $e^-p$ data~\cite{zfp:c65:627}.

\section{Phenomenological model}
\label{sec-the}

It is convenient to choose a specific phenomenological model to quantify the
experimental sensitivity which, for a narrow resonance, depends only on its
mass and decay angular distribution. The most commonly used
model~\cite{zfp:c29:115,pr:d42:815,zfp:c57:425} is based on the assumptions 
that the excited fermions have spin and isospin 1/2 and both left-handed, 
$F^*_L$, and
right-handed components, $F^*_R$, are in weak isodoublets. The Lagrangian
describes the transitions between known fermions, $F_L$, and excited states:
\begin{equation}\label{lagrangian}
{\mathcal{L}}_{F^{*}F}= \frac{1}{\Lambda}\, \bar{F^{*}_{R}}\, \sigma^{\mu
\nu} \left[g \,f\, \frac{\vec{\tau}}{2}\, \partial_{\mu} \vec{W}_{\nu}+g' \,f'\,
\frac{Y}{2}\, \partial_{\mu} B_{\nu } + g_s \,f_s\, \frac{\lambda^a}{2}\,
\partial_{\mu} G_{\nu }^{a} \, \right]F_{L}+ {\rm h.c.}\, ,
\end{equation}
where $\Lambda$ is the compositeness scale; $\vec{W}_{\nu}$, $B_{\nu}$ and
$G_{\nu}^{a}$ are the $SU(2)$, $U(1)$ and $SU(3)$ fields; $\vec{\tau}$, $Y$ and
$\lambda^a$ are the corresponding gauge-group generators; and $g$, $g'$ and
$g_s$ are the coupling constants. The free parameters $f$, $f'$ and $f_s$ are
weight factors associated with the three gauge groups and depend on the specific
dynamics describing the compositeness. For an excited fermion to be observable,
$\Lambda$ must be finite and at least one of $f$, $f'$ and $f_s$ must be
non-zero. By assuming relations between $f$, $f'$ and $f_s$, the branching
ratios of the excited-fermion decays can be fixed, and the cross section depends
only on $f/\Lambda$.

%

For excited electrons, the conventional relation $f = f'$ is adopted. The
dominant contribution to $e^*$ production is $t$-channel $\gamma$ exchange, in
which roughly 50\% of the excited electrons would be produced
elastically~\cite{zfp:c29:115}.

For excited quarks, $f=f'$ is also adopted. There are stringent limits on $f_s$
in $q^*$ production from the Tevatron~\cite{pr:d55:5263}. In this paper, $f_s$
is set to zero, and the HERA sensitivity to the electroweak couplings $f$ and
$f^{\prime}$ is exploited.  Under this assumption, $q^*$ production via $qg$
fusion vanishes and $q^*$ does not decay into $qg$.  Furthermore, a single
mass-degenerate doublet $(u^{*},d^{*})$ is assumed, so that the production
cross-section arises from both $u$- and $d$-quark excitations.

Since excited-neutrino production requires $W$ exchange, the cross section for
$M_{\nu^{*}}>200\gev$ in $e^-p$ collisions is two orders of magnitude higher
than that in $e^+p$.  Therefore, $e^-p$ reactions offer much greater sensitivity
for the $\nu^*$ search than $e^+p$ reactions.  In this paper, two very 
different assumptions are contrasted: the first uses $f=f'$, so that the 
photonic decay of
the $\nu^*$ is forbidden; the second uses $f=-f'$, so that all $\nu^*$ decays
into $\nu \gamma$, $\nu Z$ and $eW$ are allowed.

\section{Experimental setup}
\label{sec-exp}

\Zdetdesc 


Charged particles were tracked in the central tracking detector (CTD)~\citeCTD,
which operates in a magnetic field of $1.43\Tesla$ provided by a thin 
superconducting coil. The CTD consists of 72~cylindrical drift-chamber 
layers, organized in 9~superlayers covering the polar-angle\footnote{ 
The ZEUS coordinate system is a right-handed Cartesian system, with the $Z$
axis pointing in the proton beam direction, referred to as the ``forward
direction'', and the $X$ axis pointing left towards the centre of HERA.
The coordinate origin is at the nominal interaction point.
The pseudorapidity is defined as $\eta=-\ln\left(\tan\frac{\theta}{2}\right)$,
where the polar angle, $\theta$, is measured with respect to the proton beam
direction. The azimuthal angle is denoted by $\phi$.}
region 
\mbox{$15^\circ<\theta<164^\circ$}. The transverse-momentum resolution for
full-length tracks is $\sigma(p_T)/p_T=0.0058p_T\oplus0.0065\oplus0.0014/p_T$,
with $p_T$ in $\Gev$.


The high-resolution uranium--scintillator calorimeter (CAL)~\citeCAL consists 
of three parts: the forward (FCAL), the barrel (BCAL) and the rear (RCAL)
calorimeters. The calorimeters are subdivided transversely into towers,
each of which subtends solid angles ranging from $0.006$ to $0.04$
steradians. Each tower is longitudinally segmented into one
electromagnetic (EMC) section and either one (in RCAL)
or two (in BCAL and FCAL) hadronic sections (HAC). 
Each HAC section consists of a single cell, while the EMC
section of each tower is further subdivided transversely into four
cells (two in RCAL). 
The CAL energy resolutions, as measured under
test-beam conditions, are $\sigma(E)/E=0.18/\sqrt{E}$ for electrons and
$\sigma(E)/E=0.35/\sqrt{E}$ for hadrons ($E$ in $\Gev$). The
arrival time of CAL energy deposits is measured with sub-nanosecond
resolution for energy deposits above $4.5~\Gev$, allowing the rejection
of non-$ep$ background.

The luminosity was measured using the Bethe-Heitler reaction $ep \rightarrow
ep\gamma$~\cite{desy-92-066,*zfp:c63:391,*acpp:b32:2025}. The resulting small-angle
energetic photons were measured by the luminosity monitor, a lead-scintillator
calorimeter placed in the HERA tunnel at $Z=-107\met$.

A three-level trigger was used to select events online. The trigger criteria
rely primarily on the energies deposited in the calorimeter. Timing cuts were
used to reject beam-gas interactions and cosmic rays.

\section{Monte Carlo simulation}
\label{sec-mc}

The Monte Carlo (MC) event generator HEXF~\cite{lsuhe-145-1993},
based on the model of Hagiwara et al.~\cite{zfp:c29:115}, was used to simulate
the excited-fermion signals. Initial-state radiation from the beam electron is
included using the Weizs\"{a}cker-Williams approximation~\cite{prep:146:1}, and
the hadronic final state is simulated using the matrix-element and parton-shower
(MEPS) model of LEPTO 6.1~\cite{cpc:101:108} for the QCD cascade and JETSET
7.4~\cite{cpc:39:347,*cpc:43:367,*cpc:82:74} for the hadronisation.

The program DJANGO6~2.4~\cite{cpc:81:381,*spi:www:django6} was used to simulate
backgrounds from neutral and charged current deep inelastic scattering (NC and
CC DIS). The hadronic final state was simulated using the colour-dipole model as
implemented in ARIADNE~4.08~\cite{cpc:71:15} for the QCD cascade. The MEPS model
was used to evaluate systematic uncertainties (see \Sect{sys}).  Backgrounds
from elastic and quasi-elastic QED-Compton scattering were simulated using
COMPTON~2.0~\cite{proc:hera:1991:1468}. Resolved and direct photoproduction
(PHP) backgrounds were simulated with the HERWIG~5.9~\cite{cpc:67:465}
generator.  PYTHIA~5.7~\cite{cpc:46:43,*manual:cern-th-6488/92} was used to
simulate backgrounds from the photoproduction of prompt photons. The
EPVEC~1.0~\cite{np:b375:3} program was used to simulate $W$~production.

All simulated events were passed through a detector simulation based on
GEANT~3.13~\cite{tech:cern-dd-ee-84-1} and were processed with the same
reconstruction and analysis programs as used for the data.

\section{Event selection}
\label{sec-event}

The selection used the following kinematic variables and particle-identification
criteria:
\begin{itemize}
\item the scalar sum of the transverse energy deposited in the CAL,
  $E_T$;
  
\item the vector sum of the transverse energy deposited in the CAL
  (missing transverse momentum), $\ptmiss$;
  
\item the difference between the total energy and the longitudinal momentum
  deposited in the CAL, $\delta=\sum_i E_i(1-\cos\theta_i)$, where the
  energies of individual CAL cells are denoted by $E_i$ and the angles
  $\theta_i$ are estimated from the geometric cell centres and the event vertex.
  For final states where no energy is lost through the rear beam-hole, the
  nominal value of $\delta$ should equal twice the electron-beam energy
  ($2E_e=55\gev$);
  
\item an identified electromagnetic (EM) cluster, which was required to have a
  minimum transverse energy ($E^{\rm EM}_T$) of $10\gev$ and a polar angle of
  $\theta_{\rm EM} <2\rad$. If the polar angle of the cluster was less than
  $0.3\rad$, the threshold was raised to $E^{\rm EM}_T>30\gev$.  An
  electromagnetic cluster was called ``isolated'' if the sum of the CAL energy
  not associated with this cluster but within an $\eta -\phi$ cone of radius 0.8
  centered on the cluster was less than $2\gev$;

  \begin{itemize}  
  \item an EM cluster was identified as a photon candidate if no track measured
    by the CTD extrapolated to within $50$ cm of the cluster;
  \item an EM cluster was identified as an electron candidate if it had a track
    with a momentum greater than $5\gev$ that extrapolated to within 10 cm of
    the cluster. If its polar angle was less than $0.3\rad$, the cluster was not
    required to have a matching track; such clusters may also be photon
    candidates;
  \end{itemize}  
  
\item the following variables were calculated using CAL cells but excluding 
  those with polar angles below $10^\circ$, to avoid a contribution from the
  proton remnant:
  \begin{itemize}  
  \item the total invariant mass, $M$;
  \item the hadronic invariant mass, $M^{\rm had}$, and transverse energy,
    $E_T^{\rm had}$, calculated excluding those CAL cells belonging to 
    electron or photon candidates;
  \item a second missing-transverse-momentum variable,
    $\ptmiss(\theta>10^\circ)$.
  \end{itemize}  
  
\end{itemize}

To reduce the non-$ep$ background, the reconstructed $Z$ position of the
interaction vertex was required to be within $\pm50\cm$ of the nominal
interaction point. In addition, pattern-recognition algorithms were used to
suppress non-$ep$ backgrounds such as cosmic rays and beam-halo muons.

In the following, the selection criteria~\cite{zfp:c76:631,zfp:c65:627} used for
the different decay modes are listed. These criteria were obtained from MC
studies of signals and backgrounds with the goal of optimising sensitivities.
\tab{bg} contains an overview of the decays.

\subsection{Search for ${\bf e^*}$ production}
\label{sec-analy}

The criteria used to select excited-electron candidates decaying into
each of the four final states listed below are as follows:

a) $e^*\rightarrow e \gamma$:
\begin{itemize}
\item two isolated EM clusters, EM1 and EM2, each with $E_T^{\rm EM}>30\gev$;
  \vspace{-0.2cm}
\item if both clusters are within the CTD acceptance, then one and only one of
  them was required to have a matching track; \vspace{-0.2cm}
\item
$35 < \delta < 65\gev$;
\vspace{-0.2cm}
\item
$\theta_{\rm EM 1}+\theta_{\rm EM 2} < 2.5\rad$.
\end{itemize}

b) $e^*\rightarrow \nu W \rightarrow \nu q' \bar q$:
\begin{itemize}
\item
$\ptmiss > 25\gev$ and $\ptmiss(\theta >10^\circ) > 20 \gev$;
\vspace{-0.2cm}
\item
$10<\delta<50\gev$;
\vspace{-0.2cm}
\item
either $E_T^{\rm had}>50\gev$ and $M^{\rm had}>60\gev$, or $E_T^{\rm had}>80\gev$ 
and $M^{\rm had}>40\gev$;
\vspace{-0.2cm}
\item
events with an isolated electron were rejected.
\end{itemize}

c) $e^*\rightarrow e Z \rightarrow e q\bar{q}$:
\begin{itemize}
\item an electron\footnote{Since, in this channel, the electron can be close to
    a hadronic jet when the $e^*$ mass is close to that of the $Z$, the electron
    was not required to be isolated.} with $E^{\rm EM}_T>25\gev$;
  \vspace{-0.2cm}
\item
$35 < \delta < 65\gev$;
\vspace{-0.2cm}
\item either $E_T^{\rm had}>60\gev$ and $M^{\rm had}>80\gev$, or $E_T^{\rm
    had}>80\gev$ and $M^{\rm had}>40\gev$; \vspace{-0.2cm}
\item 
  $0.8<M/M_{eZ}<1.2$, where $M_{e Z}$ is the electron-$Z$ invariant mass
  defined in \Sect{mass}.
\end{itemize}

d) $e^*\rightarrow eZ \rightarrow e\nu\bar{\nu}$:
\begin{itemize}
\item
an isolated electron;
\vspace{-0.2cm}
\item
$\ptmiss> 20\gev$;
\vspace{-0.2cm}
\item $\cos(\phi_{e}-\phi_{\rm had}) > -0.95$ if $E_T^{\rm had} >2\gev$, where
  $\phi_{e}$ and $\phi_{\rm had}$ are the azimuthal angles of the electron and
  the hadronic system\footnote{This cut rejects background from NC DIS events
    where the hadronic system balances the scattered electron back-to-back in
    $\phi$.}, respectively; \vspace{-0.2cm}
\item
$\ptmiss/E_T > 0.4$.
\end{itemize}

\subsection{Search for ${\bf q^*}$ production}

The following selection criteria were used for the excited-quark search:

a) $q^*\rightarrow q \gamma$:
\begin{itemize}
\item
an isolated photon with $E_T^{\rm EM}>20\gev$ and $\theta_{\rm EM}<1.2\rad$;
\vspace{-0.2cm}
\item
$E_T^{\rm had}>40\gev$.
\end{itemize}

b) $q^*\rightarrow q W \rightarrow q e \nu$:
\begin{itemize}
\item
an isolated electron with $E_T^{\rm EM}>15\gev$;
\vspace{-0.2cm}
\item
$\ptmiss>18\gev$;
\vspace{-0.2cm}
\item
$\cos(\phi_{e} - \phi_{\rm had}) >-0.95$;
\vspace{-0.2cm}
\item
$E_T^{\rm had}>5\gev$.
\end{itemize}

\subsection{Search for ${\bf \nu^*}$ production}

The following cuts were applied to select excited neutrinos decaying into the
three final states listed below:

a) $\nu^*\rightarrow \nu \gamma$:
\begin{itemize}
\item
an isolated photon with $E_T^{\rm EM}>20\gev$;
\vspace{-0.2cm}
\item
$\ptmiss >25\gev$;
\vspace{-0.2cm}
\item
$E_T > 50\gev$;
\vspace{-0.2cm}
\item
$\delta<45\gev$.
\end{itemize}

b) $\nu^*\rightarrow \nu Z \rightarrow \nu q\bar q$: 
\begin{itemize}
\item the same cuts as used for $e^*\rightarrow \nu W \rightarrow \nu q' \bar q$
  were applied.
\end{itemize}
  
c) $\nu^*\rightarrow e W \rightarrow e q'\bar q$:
\begin{itemize}
\item the same cuts as used for $e^*\rightarrow e Z\rightarrow e q\bar q$ were
  applied, except that the cut on $M/M_{eZ}$ was replaced by the cut
  $0.9<M/M_{eW}<1.2$, where $M_{eW}$ is the electron-$W$ invariant mass defined
  in the next section.
\end{itemize}

\section{Mass reconstruction of excited fermions}
\label{sec-mass}

To improve the mass resolution, three kinematic constraints could be applied:
\begin{itemize}
\item the transverse momentum of the excited fermion was assumed to be zero.
  This was used except for $\nu^*\to\nu\gamma$;
\item the longitudinal-momentum variable, $\delta$, of the excited fermion's
  decay products was set to twice the electron-beam energy, the value 
  expected when all decay products are
  observed. This assumption is less justified for decays of $\nu^*$ and $q^*$
  than for $e^*$, leading to worse resolutions for $\nu^*$ and $q^*$.
  Therefore, in the two cases $\nu^*\to\nu\gamma$ and $q^*\to q\gamma$, this
  constraint was not used;
\item in all decays involving a final-state $W$ or $Z$, the mass of their decay
  products was constrained to be the mass of the respective boson.
\end{itemize}

                               

For $e^*\rightarrow e \gamma$, the electron-photon invariant mass was determined
by the double-angle method~\cite{proc:hera:1991:23,*proc:hera:1991:43} as
\begin{eqnarray*}
M^{2}_{e\gamma}=(2E_{e})^{2}\left(
\frac{\sin\theta_{\gamma}}{1-\cos\theta_{\gamma}}\right)
\left(
\frac{\sin\theta_{e}}{1-\cos\theta_{e}}\right)\, ,
\end{eqnarray*}
where $\theta_{e}$ and $\theta_{\gamma}$ are the polar angles of the
electron and photon, respectively.

For $q^*\rightarrow q\gamma$, the $q\gamma$ invariant mass was obtained from
\begin{equation*}
M_{q\gamma}^2=2E_{\gamma}^2\frac{\sin \theta_{\gamma}}{\sin
  \theta_{\rm had}}\Big[1-\cos(\theta_{\gamma}+\theta_{\rm had})\Big]\, ,
\end{equation*}
where $E_{\gamma} $ and $\theta_{\rm had}$ are the
energy of the photon and the polar angle of the hadronic
system, respectively.

For $\nu^*\rightarrow \nu \gamma$, the mass of the excited neutrino was
determined from the invariant mass of the photon and the neutrino.  The
four-momentum of the neutrino was obtained using energy-momentum conservation.

For the excited-fermion decays to a fermion and a heavy vector boson, $F^* \to
FV$, the mass was reconstructed using the energy and longitudinal momentum of
the two decay products:
\begin{equation*}
M^2_{FV} = 2 E_e (E^F + p_Z^F + E^V + p_Z^V)\, .
\end{equation*}
By using the relation
\begin{equation*} 
(E^V - p_Z^V) (E^V + p_Z^V) = M_V^2 + (E^F - p_Z^F) (E^F + p_Z^F)\, ,
\end{equation*}
the formula can be written as
\begin{equation*}  
M^2_{FV} = 2 E_e \frac{2E_e (E^F + p_Z^F) + M_V^2}{2E_e - (E^F - p_Z^F)}\, .
\end{equation*}
 
For $e^*\to eZ\to eq\bar q$, $e^*\to eZ\to e\nu\bar\nu$ and $\nu^*\to eW\to
eq'\bar q$, the final-state electron energy and polar angle were used to obtain
$E^F$ and $p_Z^F$.
 
For $q^*\to qW\to qe\nu$, $E^F$ and $p_Z^F$ were obtained using the CAL cells
with polar angle $\theta > 10^\circ$, excluding those belonging to the
electron.
 
For $e^*\to\nu W\to\nu q'\bar q$ and $\nu^*\to\nu Z\to\nu q\bar q$, the neutrino
variables were obtained from the hadronic system using the relations $E^F -
p_Z^F = 2E_e - \delta$ and $E^F + p_Z^F = \ptmiss^2 / (2E_e - \delta)$.
 
The Gaussian mass resolutions and the overall efficiencies after all selection
cuts are listed in \tab{res} for excited fermions with masses of $125$ and
$250\gev$.

\section{Systematic uncertainties}
\label{sec-sys}

The most important sources of systematic uncertainty were:

\begin{itemize}
\item the theoretical uncertainty on the production cross-section due to
  radiative corrections to the excited-fermion production model and to the
  uncertainties on the parton densities in the proton was taken to be 8\%, as
  determined from an earlier study \cite{zfp:c65:627};
\item the acceptance was determined using a simulation of spin-1/2 excited
  fermions. To estimate the effect of models assuming other spin states, the
  variation of the acceptance was evaluated by changing the nominal decay-angle
  distribution\footnote{For $F^{*} \rightarrow F \gamma$, for example, the
    nominal distribution is ($1+\cos \theta ^{*}$), where $\theta ^{*}$ is the
    polar angle between the incoming and outgoing fermions in the $F^{*}$ rest
    frame.} to an isotropic one~\cite{zfp:c65:627}.  The variation was typically
  5\% or less;
\item the energy scale of the calorimeter was varied by $\pm3\%$, leading to
  uncertainties in the excited-fermion efficiency of at most 3\%;
\item the uncertainty on the measured integrated luminosity of the 1994--1997
  $e^+p$ data sample was 1.6\% and that of the 1998--1999 $e^-p$ data sample
  was 1.8\%.
\end{itemize}

\newcommand{\like}{{\cal L}}
\newcommand{\mtru}{{\tilde M}}
\newcommand{\signwa}{\sigma_{\rm nwa}}
\newcommand{\Mf}{M_{F^*}}
\section{Results}
\label{sec-res}

\begin{figure}[htbp]
  \begin{center}
    {\bf\LARGE ZEUS}\\
    \epsfig{file=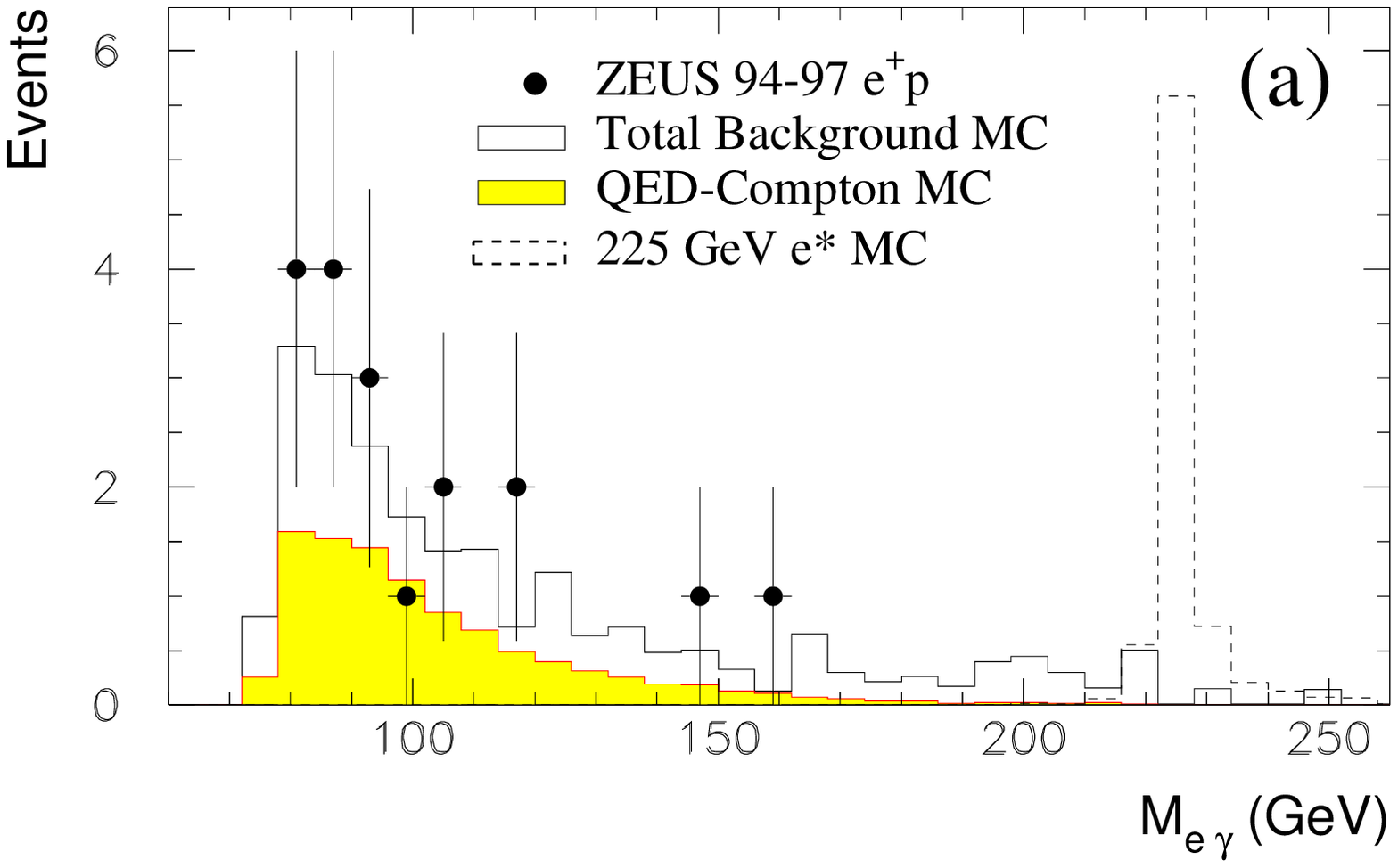,width=100mm}\\[5mm]
    \epsfig{file=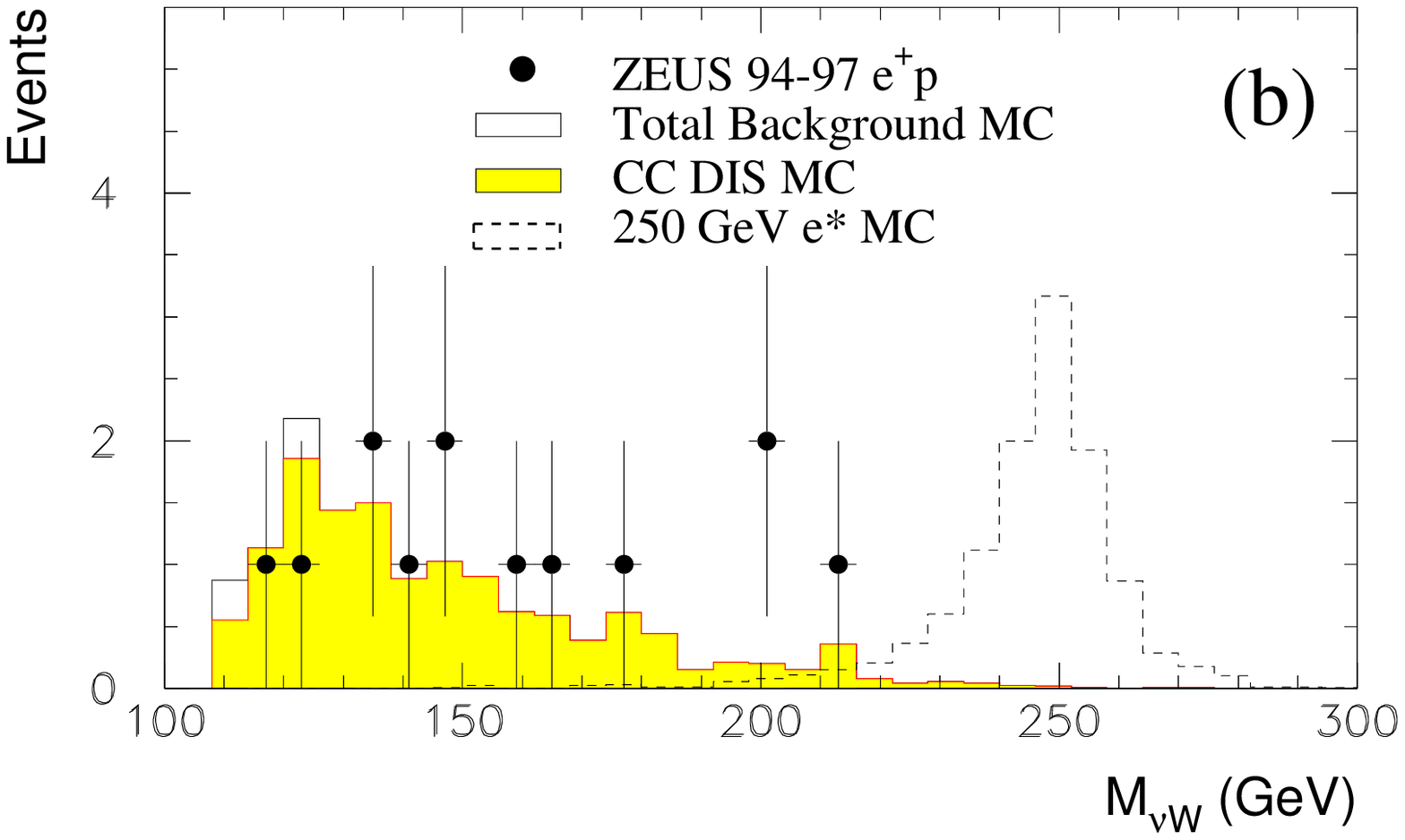,width=100mm}\\[5mm]
    \epsfig{file=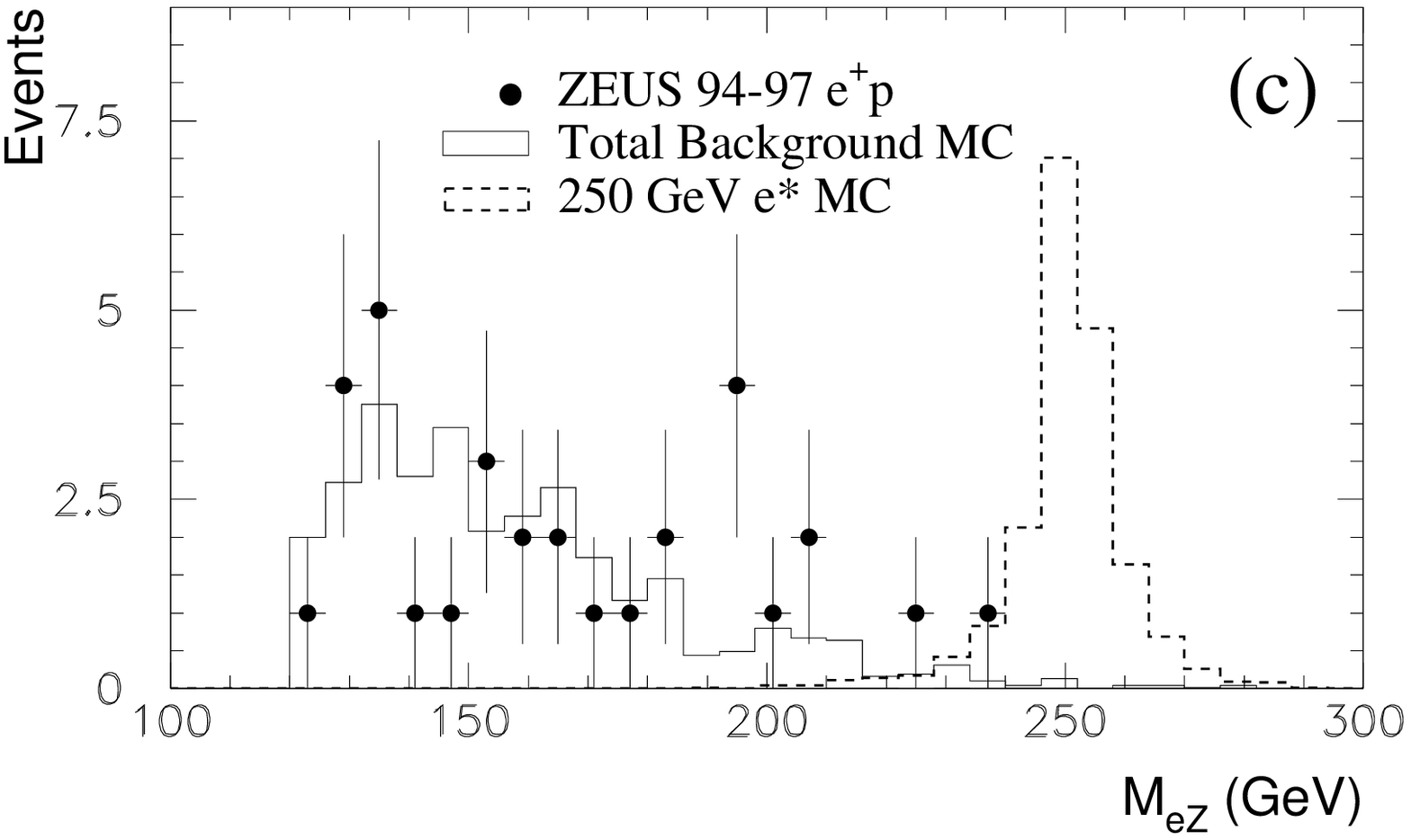,width=100mm}
    \caption{Invariant-mass distributions for
      (a) $e^*\rightarrow e\gamma$, (b) $e^*\rightarrow \nu W\rightarrow \nu
      q'\bar q$ and (c) $e^*\rightarrow e Z\rightarrow e q\bar q$. Examples of
      $e^*$ signals are shown as the dashed histograms (arbitrary
      normalisation) to illustrate the mass resolution.}
    \label{fig-estar_mass}
  \end{center}
\end{figure}

\begin{figure}[htbp]
  \begin{center}
    {\bf\LARGE ZEUS}\\
    \epsfig{file=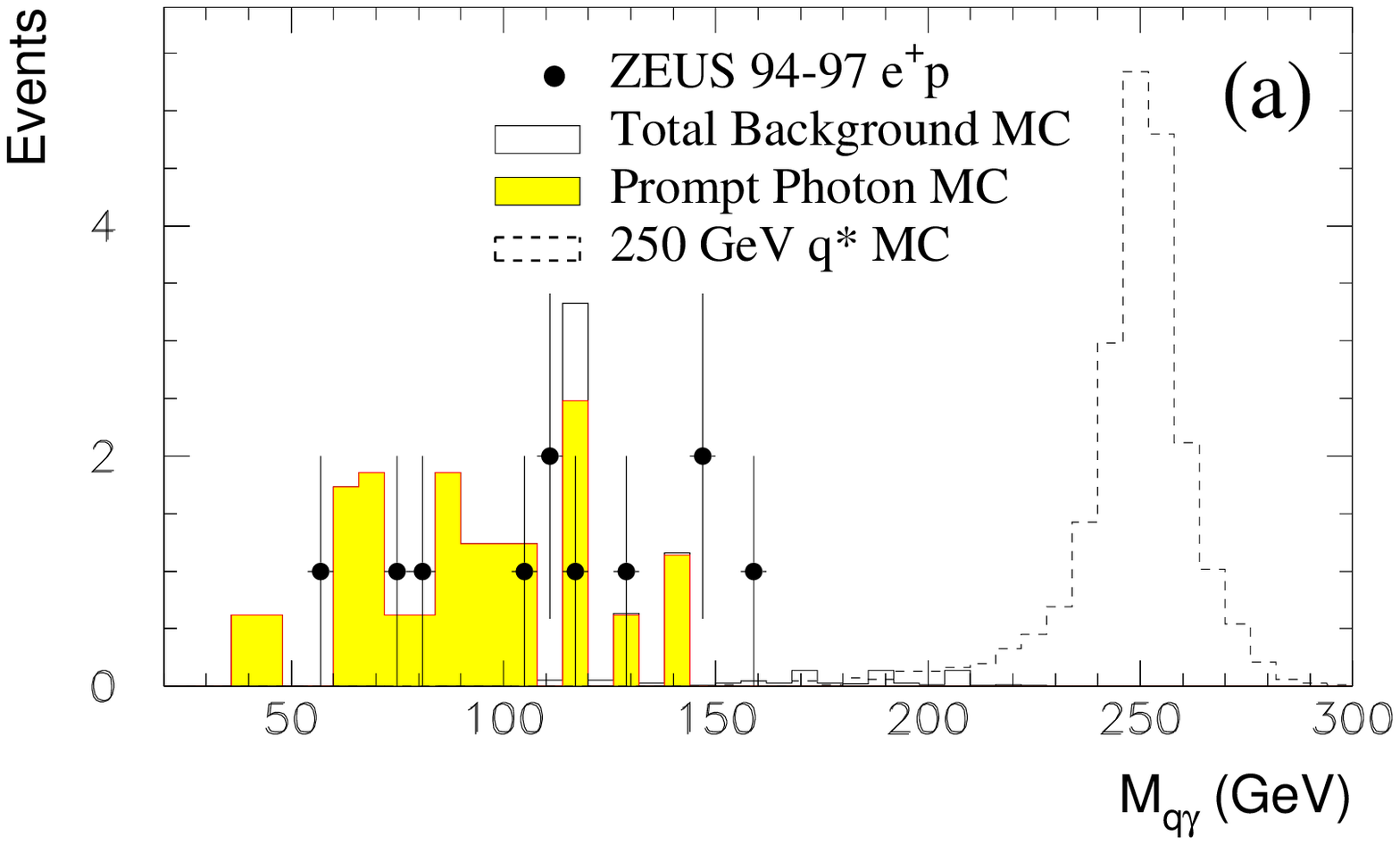,width=100mm}\\[5mm]
    \epsfig{file=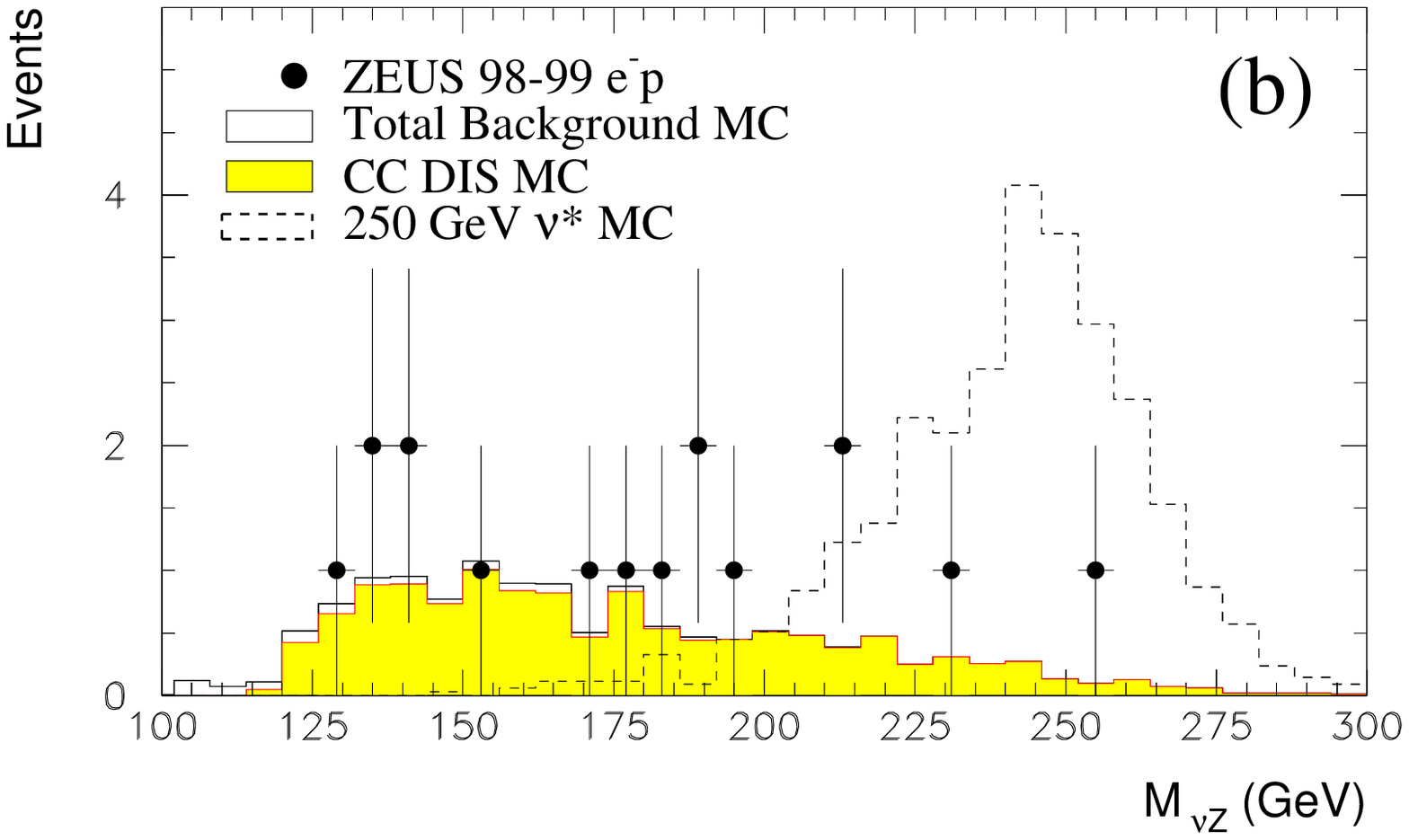,width=100mm}\\[5mm]
    \epsfig{file=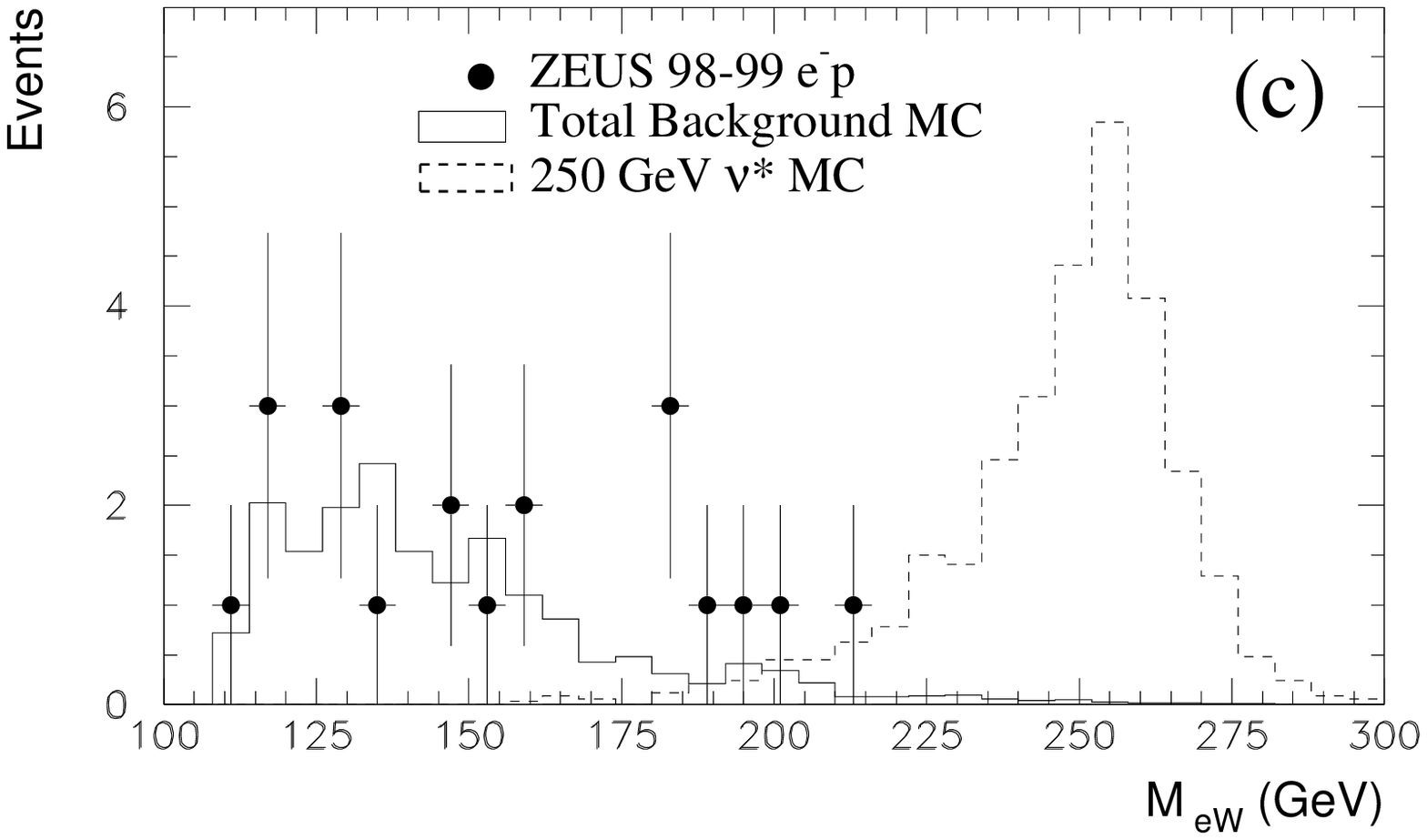,width=100mm}
    \caption{Invariant-mass distributions for
      (a) $q^*\to q\gamma$, (b) $\nu^*\rightarrow \nu Z\to\nu q\bar q$ and (c)
      $\nu^*\rightarrow e W\to eq'\bar q$. Examples of $q^*$ and $\nu^*$
      signals are shown as the dashed histograms (arbitrary
      normalisation) to illustrate the mass resolution.}
   \label{fig-qstar-nustar_mass}
  \end{center}
\end{figure}

The number of observed events and the expected background for each channel are
shown in \tab{bg}. No significant excess of events is observed.  The
distributions of the invariant mass
are compared\footnote{Only channels with more than ten
  candidate events are shown.} 
in Figs.~\ref{fig-estar_mass} and
\ref{fig-qstar-nustar_mass} with the expected backgrounds for $e^*$, $q^*$ and
$\nu^*$. No evidence for a resonance is seen.

Since there is no evidence for excited fermions, upper limits at 95\% confidence
level on $f/\Lambda$ were derived. A Bayesian
technique with the prior flat in $(f/\Lambda)^2$ was used.
The limit on $\xi=f/\Lambda$ was given as the solution to
$0.95\int_0^\infty d\xi^2 \like=\int_0^{\xi^2_{\rm lim}} d\xi^2\like$
where 
\[
\like(\xi)= \int_0^\infty d\gamma 
{1\over{\sqrt{2\pi}\sigma_\gamma}}
\exp{{-(\gamma-1)^2}\over{2\sigma_\gamma^2}}
\prod_c \left\{
p\left(N_c,\gamma(S_c(\xi)+B_c)\right)
\prod_{i=1}^{N_c} \left({{s_c(M_{ic},\xi)+b_c(M_{ic})}\over{S_c(\xi)+B_c}}\right)
\right\}.
\]
Here $c$ labels the decay channel, $N_c$ denotes the number of events         
observed in that channel and
$M_{ic}$ is the reconstructed mass of the $i$-th observed event.
The probability to observe $n$ events in a Poisson
process with mean $\lambda$ is denoted $p(n,\lambda)$.
The expected reconstructed mass spectra for signal
and background in channel $c$ are denoted by $s_c$ and $b_c$ respectively.
The number of expected signal events is given by $S_c(\xi)=\int dM s_c(M,\xi)$
and the expected background is given by $B_c=\int dM b_c(M)$.
Systematic uncertainties were taken into account by integration over $\gamma$,
which has a Gaussian distribution with mean 1 and width $\sigma_\gamma=0.123$.
The systematic uncertainties degrade the limits by at most 4\%.

The spectra of reconstructed mass, $M$, for the signals is calculated as
\[s_c(M,\xi)=L\int_0^\infty d\mtru \signwa(\mtru,\xi) 
{{2\mtru\Mf\Gamma}\over{\left(\mtru^2-\Mf^2\right)^2+\left(\Mf\Gamma\right)^2}}
\beta_c(\mtru)\varepsilon_c(\mtru) D_c(M,\mtru)\]
where $L$ is the integrated luminosity,
$\mtru$ denotes the true mass of the produced excited fermion and $\Mf$ and
$\Gamma$ are the pole mass and the width.
The total cross section in the narrow-width approximation is denoted $\signwa$.
The functions $\beta_c(\mtru)$ and $\varepsilon_c(\mtru)$ respectively label
the branching ratio and
the detector acceptance for decay channel $c$.
Detector resolution was described
by the function $D_c(M,\mtru)$ which gives the probability density in
$M$ for events with true mass $\mtru$, as obtained
by a fit to simulated signal events. The function $D_c(M,\mtru)$ is a linear
combination of a Gaussian in $M$ with mean $M_0$ and a function of the form
$\left[(\exp(\alpha(M-M_0))+\exp(-\beta(M-M_0))\right]^{-1}$
with $\alpha$ and $\beta$
positive. The latter function accounts for the tails of the mass spectrum.

\begin{figure}[htbp]
\begin{center}
{\bf\LARGE ZEUS}\\[5mm]
    \epsfig{file=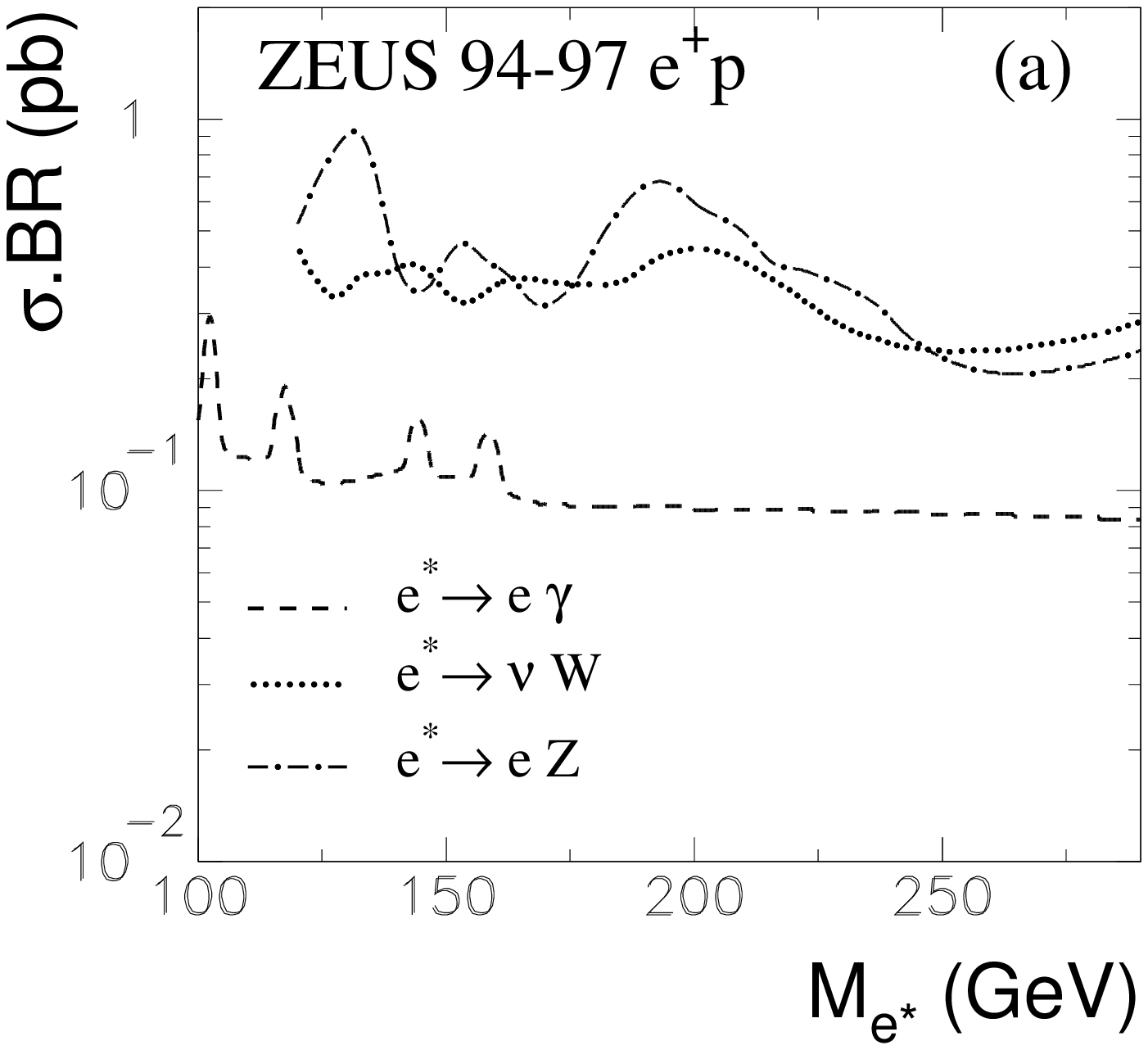,width=0.49\textwidth}\hfill
    \epsfig{file=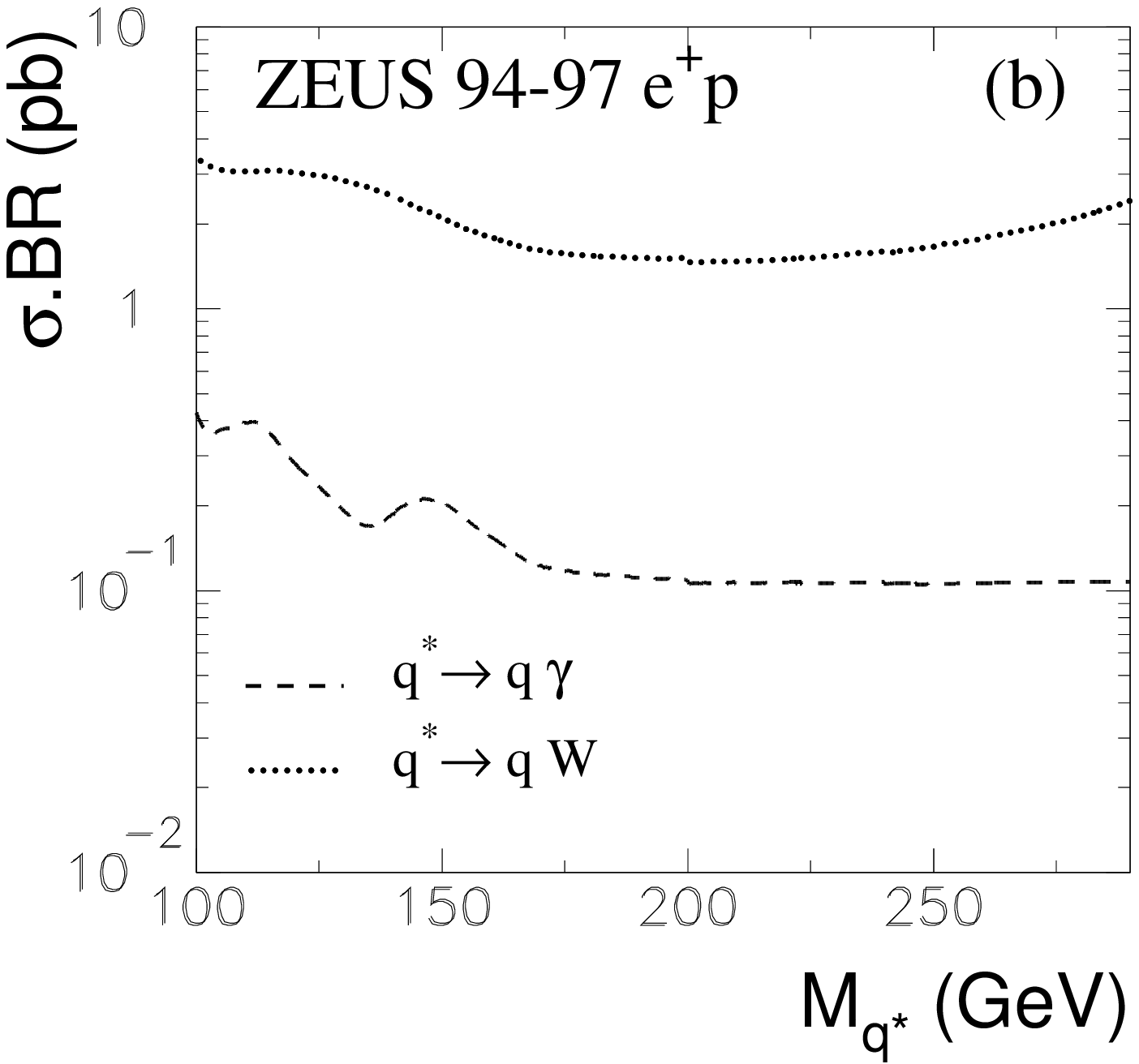,width=0.49\textwidth}\\[2mm]
    \epsfig{file=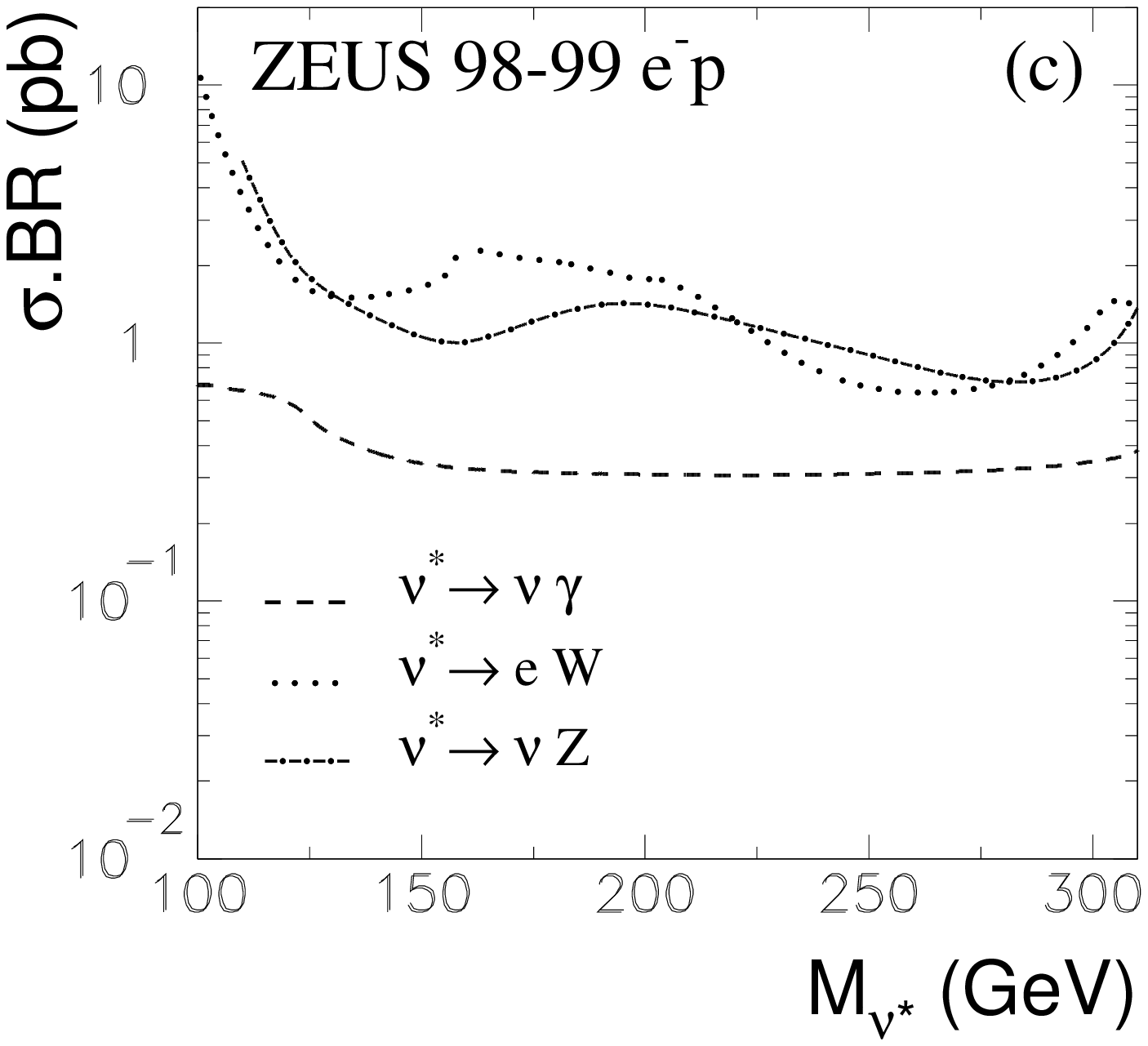,width=0.49\textwidth}
    \caption{Upper limits at $95\%$ confidence level on the production cross section times the
      branching ratio as a function of the excited-fermion mass for (a)
      $e^*\rightarrow e\gamma$, $e^*\rightarrow \nu W$, $e^*\rightarrow eZ$, (b)
      $q^*\rightarrow q\gamma$, $q^*\rightarrow q W$ and (c) $\nu^*\rightarrow
      \nu\gamma$, $\nu^*\rightarrow e W$, $\nu^*\rightarrow \nu Z$.  In all
      cases, the areas above the lines are excluded.}
    \label{fig-lim-sigbr}
\end{center}
\end{figure}

\Fig{lim-sigbr} shows limits on $\sigma \times BR$ under the assumption
of a vanishing width. In this case $s_c(M,\xi)$ reduces to
$L\signwa(\Mf,\xi) \beta_c(\Mf)\varepsilon_c(\Mf) D_c(M,\Mf)$.
The limits were
obtained using ARIADNE, the hadronisation model that gives the background
estimate leading to the more conservative limits. The alternative choice of
MEPS would have resulted in limits up to 12.5\% more stringent.

\begin{figure}[htbp]
\begin{center}
{\bf\LARGE ZEUS}\\[14mm]
  \epsfig{file=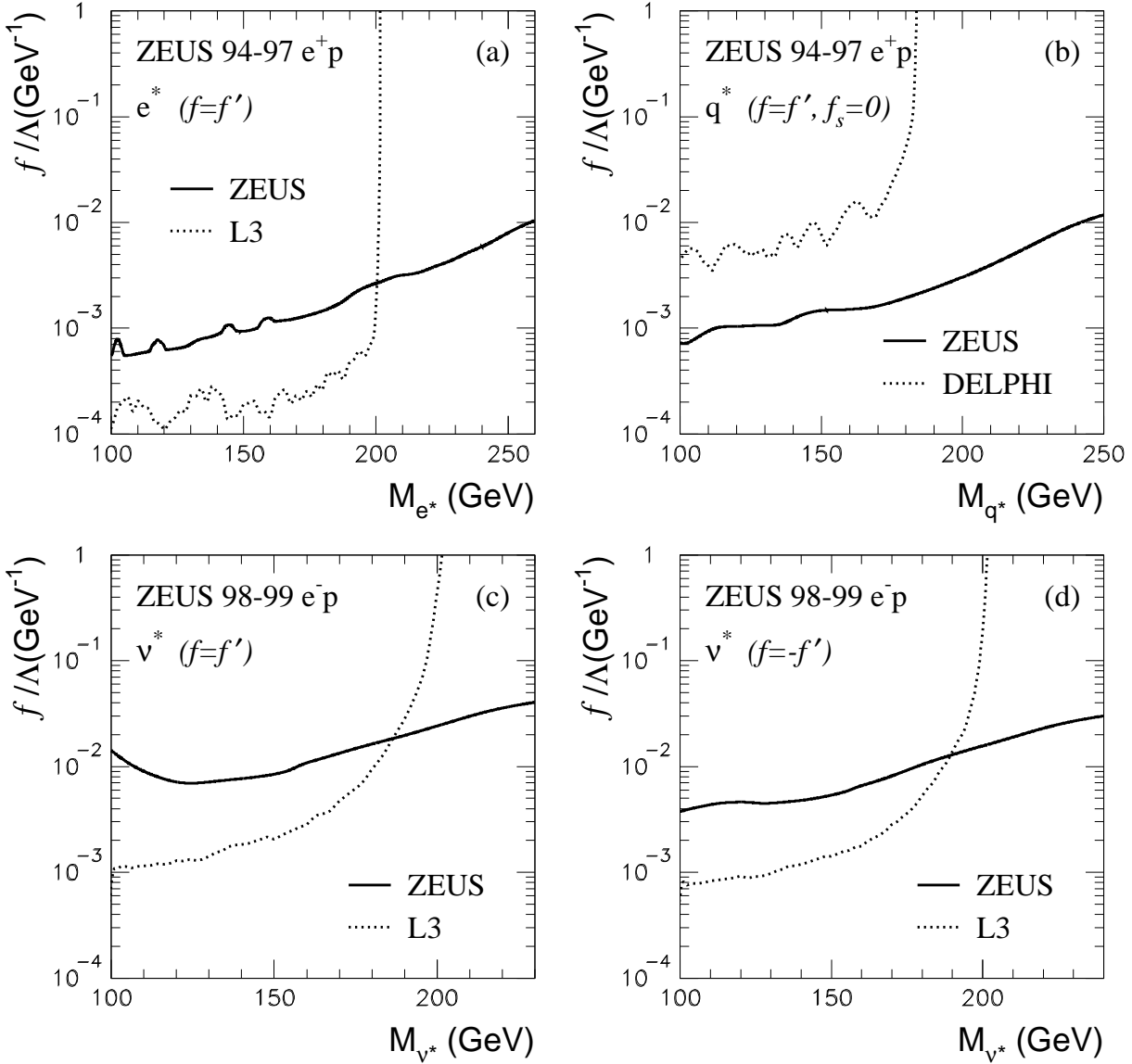,width=0.97\textwidth,bb=30 30 537 497}
  \caption{Upper limits at $95\%$ confidence level on
    the coupling $f/\Lambda$ as a function of the excited-fermion mass for (a)
    excited electrons assuming $f=f'$, (b) excited quarks assuming $f=f'$, $f_s
    = 0$, (c) excited neutrinos assuming $f= f'$ and (d) excited neutrinos
    assuming $f=-f'$.  The solid curves result from combining all channels.
    The dotted lines are the limits from L3~\protect\cite{pl:b502:37} and
    DELPHI~\protect\cite{epj:c8:41}. The DELPHI limit on $q^*$ was derived
    assuming BR$(q^*\to q\gamma)=1$.  In all cases, the areas above the lines
    are excluded.  }
    \label{fig-lim-flambda}
\end{center}
\end{figure}

Next, using the model~\cite{zfp:c29:115,pr:d42:815,zfp:c57:425} discussed
in \Sect{the} to calculate the natural width $\Gamma$
and the method described above, 95\% confidence-level
upper limits on $f/\Lambda$ as a function of mass were calculated
for excited electrons, quarks and neutrinos, shown in \fig{lim-flambda}.

By assuming $f/\Lambda=1/M_{F^*}$ and $f=f'$, excited fermions were
excluded in the mass intervals from $100$ up to $228\gev$ for $e^*$,
$205\gev$ for $q^*$, and $135\gev$ for $\nu^*$. 
For $f=-f'$, excited neutrinos were excluded up to $158\gev$.

The exclusion limits on $f/\Lambda$ are compared with
corresponding direct limits from LEP experiments~\cite{pl:b502:37,epj:c8:41}
in~\fig{lim-flambda}. The
corresponding H1 limits on $e^*$ and $q^*$~\cite{epj:c17:567} are comparable to
those presented here. As seen in~\fig{lim-flambda}, the ZEUS limits extend to
significantly higher masses than those of LEP. The present limits on
$f/\Lambda$ for $q^*$ constrain the electroweak $q^*$ couplings and are
complementary to the limits on $f_s$ set by CDF~\cite{pr:d55:5263},
which constrain the strong $q^*$ coupling.

 
\section{Conclusions}
\label{sec-conc}

A search for heavy excited electrons which decay into $e\gamma$, $eZ$ or $\nu W$
has been performed using $47.7\pbi$ of $e^{+}p$ data at a centre-of-mass energy
of $300\gev$ collected with the ZEUS detector at HERA during 1994 to 1997. There
is no evidence for a narrow resonance decaying to any of the final states
considered here.  Upper limits at 95\% confidence level on $\sigma\times BR$ and
$f/\Lambda$ have been derived. Assuming $f/\Lambda = 1/M_{e^*}$ and $f=f'$,
excited electrons are excluded in the mass range $100\gev$ to $228\gev$.

The same data sample has been used to search for heavy excited quarks decaying
to $q\gamma$ or $qW$. No evidence is found for such resonances, so that
exclusion limits have been set. Assuming $f/\Lambda = 1/M_{q^*}$, $f=f'$ and
$f_s=0$, excited quarks are excluded in the mass range $100\gev$ to $205\gev$.

A search for heavy excited neutrinos which decay into $\nu\gamma$, $\nu Z$ or $e
W$ has been performed using $16.7\pbi$ of $e^{-}p$ data at a centre-of-mass
energy of $318\gev$ collected in 1998 and 1999. No resonance has been observed,
and upper limits on $\sigma\times BR$ and $f/\Lambda$ have been derived.
Assuming $f/\Lambda = 1/M_{\nu^*}$ and $f=f'$ ($f=-f'$), excited neutrinos are
excluded in the mass range $100\gev$ to $135\;(158)\gev$.

\section*{Acknowledgements}
\label{sec-ack}

We appreciate the contributions to the construction and maintenance of the ZEUS
detector by many people who are not listed as authors. We thank the HERA machine
group for their outstanding operation of the collider. The support of the DESY
computing and network services is gratefully acknowledged. Finally, we thank the
DESY directorate for their strong support and encouragement.

\vfill\eject

{
\def\bibname{\Large\bf References}
\def\refname{\Large\bf References}
\pagestyle{plain}
\ifzeusbst
  \bibliographystyle{../BiBTeX/bst/l4z_default}
\fi
\ifzdrftbst
  \bibliographystyle{../BiBTeX/bst/l4z_draft}
\fi
\ifzbstepj
  \bibliographystyle{../BiBTeX/bst/l4z_epj}
\fi
\ifzbstnp
  \bibliographystyle{../BiBTeX/bst/l4z_np}
\fi
\ifzbstpl
  \bibliographystyle{../BiBTeX/bst/l4z_pl}
\fi
{\raggedright
\bibliography{../BiBTeX/user/syn.bib,%
              ../BiBTeX/bib/l4z_articles.bib,%
              ../BiBTeX/bib/l4z_books.bib,%
              ../BiBTeX/bib/l4z_conferences.bib,%
              ../BiBTeX/bib/l4z_h1.bib,%
              ../BiBTeX/bib/l4z_misc.bib,%
              ../BiBTeX/bib/l4z_old.bib,%
              ../BiBTeX/bib/l4z_preprints.bib,%
              ../BiBTeX/bib/l4z_replaced.bib,%
              ../BiBTeX/bib/l4z_temporary.bib,%
              ../BiBTeX/bib/l4z_zeus.bib}}
}
\vfill\eject

\begin{table}[p]
\begin{center}
    \begin{tabular}{|l|c|c|c|c|}
      \hline
      Decay Mode & Background & Data & Predicted  \\
                 & processes  &      & background \\
      \hline\hline
      $e^*\rightarrow e\gamma$ & 
      \small NC, QED-Compton
      & $18$ & $20.1\pm 1.2$
      \\ 
      $e^*\rightarrow \nu W\rightarrow \nu q'\bar q$ &
      \small CC DIS, PHP
      & $13$ &  $13.9\pm 1.1$
      \\
      $e^*\rightarrow e Z\rightarrow eq\bar q$ &
      \small NC DIS 
      & $32$ & $32.9\pm 1.1$
      \\
      $e^*\rightarrow e Z\rightarrow e\nu \bar\nu$ &
      \small NC DIS, W, CC DIS  
      & $1$ & $4.1\pm 0.6$
      \\ \hline
      $q^*\rightarrow q\gamma$ & 
      \small Prompt $\gamma$, PHP, NC DIS
      & $11$ & $19.0\pm 1.9$
      \\
      $q^*\rightarrow qW\rightarrow qe\nu$ &
      \small NC DIS, W, CC DIS
      & $4$ & $4.1\pm 0.6$
      \\ \hline
      $\nu^*\rightarrow \nu\gamma$ & 
      \small CC DIS
      & $2$ & $1.5\pm 0.2$
      \\ 
      $\nu^*\rightarrow \nu Z\rightarrow \nu q\bar q$ &
      \small CC DIS, PHP
      & $16$ & $13.5\pm 0.6$
      \\
      $\nu^*\rightarrow e W\rightarrow eq'\bar q$ &
      \small NC DIS
      & $20$ & $15.0\pm 1.3$
      \\
      \hline
\end{tabular}
\caption{The excited-fermion decay modes, main backgrounds and numbers of events
  that pass the selection criteria for the different channels compared with the
  Monte Carlo background predictions. The abbreviations of the background
  processes are defined in the text. The $e^*$ and $q^*$ results are from
  $47.7\pbi$ of $e^+p$ data and the $\nu^*$ results are from $16.7\pbi$ of
  $e^-p$ data. The uncertainties on the background predictions are statistical
  only. }
  \label{tab-bg}
\end{center}
\end{table}


\begin{table}[p]
\begin{center}
    \begin{tabular}{|l|c|c|c|c|}
      \hline
      Decay Mode & \multicolumn{2}{c|}{Resolution ($\gev$)} & \multicolumn{2}{c|}{Efficiency ($\%$)}\\
                 & $125\gev$ & $250\gev$ & $125\gev$ & $250\gev$ \\ \hline\hline
      $e^*\rightarrow e\gamma$ & 1.0 & 2.2 &  66 & 78  \\ 
      $e^*\rightarrow \nu W\rightarrow \nu q'\bar q$ &  5.1 & 9.5 & 48 & 50 \\
      $e^*\rightarrow e Z\rightarrow eq\bar q$ & 3.3 & 6.0 & 27 & 52 \\
      $e^*\rightarrow e Z\rightarrow e\nu \bar\nu$ &  3.3 & 8.2 & 58 & 82 \\ \hline
      $q^*\rightarrow q\gamma$ & 4.7 & 9.1 & 55 & 67 \\
      $q^*\rightarrow qW\rightarrow qe\nu$ & 7.6 & 19.0 & 46 & 38 \\ \hline
      $\nu^*\rightarrow \nu\gamma$ & 5.6 & 5.3 & 55 & 61 \\ 
      $\nu^*\rightarrow \nu Z\rightarrow \nu q\bar q$ &  8.3 & 17.2 & 39 & 66 \\
      $\nu^*\rightarrow e W\rightarrow eq'\bar q$ & 14.4 & 15.3 & 49 & 51 \\
      \hline
\end{tabular} 
\caption{Gaussian mass resolutions and selection efficiencies for 
  excited fermions with masses of $125$ and $250\gev$.}
  \label{tab-res}
\end{center}
\end{table}


\vspace{2cm}

%
%
\end{document}